\documentclass[prl,aps,twocolumn]{revtex4}

\usepackage{psfrag}
\usepackage{graphicx}
\usepackage{dcolumn}
\usepackage{latexsym,amsfonts}
\usepackage{bm}
\usepackage{amssymb}
\begin{document}

\title{On Bound-State $\beta^-$--Decay Rate of the Free Neutron}

\author{M. Faber$^{a}$\thanks{E-mail: faber@kph.tuwien.ac.at},
A. N. Ivanov${^{a,b}}$, V. A. Ivanova$^{c}$, J. Marton$^{b}$,
M. Pitschmann${^a}$, N. I. Troitskaya$^{c}$, M. Wellenzohn$^{a}$}
\affiliation{${^a}$Atominstitut der \"Osterreichischen
Universit\"aten, Technische Universit\"at Wien, Wiedner Hauptstrasse
8-10, A-1040 Wien, Austria} \affiliation{${^b}$Stefan Meyer Institut
f\"ur subatomare Physik \"Osterreichische Akademie der Wissenschaften,
Boltzmanngasse 3, A-1090, Wien, Austria}\affiliation{ $^e$ State
Polytechnic University of St. Petersburg, Polytechnicheskaya 29,
195251, Russian Federation} \email{ivanov@kph.tuwien.ac.at}

\date{\today}

\begin{abstract}
We calculate the bound-state $\beta^-$--decay rate of the free
neutron. We show that hydrogen in the final state of the decay is
produced with a probability of about $99\,\%$ in the hyperfine state
with zero orbital $\ell = 0$ and atomic angular momentum $F = 0$. 
\end{abstract}

\maketitle

\section{Introduction}

The continuum-state $\beta^-$--decay of the free neutron $n \to p +
e^- + \tilde{\nu}_e$ is well measured experimentally \cite{PDG06} and
investigated theoretically \cite{EK66}--\cite{STW2}. Recently
\cite{WS1,WS2} Schott {\it et al.} have reported the experimental data
on the bound-state $\beta^-$--decay of the free neutron $n \to {\rm H}
+ \tilde{\nu}_e$. In this letter we apply the technique, which we used
for the analysis of the weak decays of the H--like, bare heavy ions
and mesic hydrogen \cite{Ivanov1}--\cite{Faber2}, to the calculation
of the bound-state $\beta^-$--decay rate of the free neutron.

\subsection{$V-A$ weak hadronic interactions}

The weak interaction Hamilton density operator we take in the form
\begin{eqnarray}\label{label1}
\hspace{-0.3in}{\cal H}_W(x) &=&
\frac{G_F}{\sqrt{2}}\,V_{ud}\,[\bar{\psi}_p(x)\gamma_{ \mu}(1 -
g_A\gamma^5)\psi_n(x)]\nonumber\\
\hspace{-0.3in}&&\times\,[\bar{\psi}_e(x)\gamma^{\mu}(1 -
\gamma^5)\psi_{\nu_e}(x)],
\end{eqnarray}
where $G_F = 1.166\times 10^{-11}\,{\rm MeV}^{-2}$ is the Fermi weak
constant, $V_{ud} = 0.97377$ is the CKM matrix element \cite{PDG06},
$g_A = 1.3$ is the axial--vector renormalisation constant and
$\psi_p(x)$, $\psi_n(x)$, $\psi_e(x)$ and $\psi_{\nu_e}(x)$ are
operators of interacting proton, neutron, electron and anti-neutrino,
respectively. The $\mathbb{T}$--matrix of weak interactions is equal
to
\begin{eqnarray}\label{label2}
\mathbb{T} = -\int d^4x\,{\cal H}_W(x).
\end{eqnarray}
In the final state of the bound-state $\beta^-$--decay hydrogen can be
produced only in the $ns$--states, where $n$ is a {\it principal}
quantum number $n = 1,2,\ldots $ \cite{Faber1,Faber2}. The
contribution of the excited $n\ell$-state with $\ell > 0$ is
negligible small. Due to hyperfine interactions \cite{HFS1} hydrogen
can be in two hyperfine states $(ns)_F$ with $F = 0$ and $F = 1$

The wave function of hydrogen ${\rm H}$ in the
$ns$--state we take in the form \cite{IV2}--\cite{IV6}
\begin{eqnarray}\label{label3}
\hspace{-0.3in}&&|{\rm H}^{(ns)}(\vec{q}\,)\rangle =
 \frac{1}{(2\pi)^3}\sqrt{2 E_{\rm H}(\vec{q}\,)}\nonumber\\
\hspace{-0.3in}&&\times\int \frac{d^3k_e}{\sqrt{2E_e(\vec{k}_e)}}
 \frac{d^3k_p}{\sqrt{2E_p(\vec{k}_p)}}\, \delta^{(3)}(\vec{q} -
 \vec{k}_e - \vec{k}_p)\nonumber\\
\hspace{-0.3in}&&\times \phi_{ns}\Big(\frac{m_p \vec{k}_e - m_e
 \vec{k}_p}{m_p + m_e}\Big) a^{\dagger}_{ns}(\vec{k}_e,\sigma_e)
 a^{\dagger}_p(\vec{k}_p,\sigma_p)|0\rangle,
\end{eqnarray}
where $E_{\rm H}(\vec{q}\,) = \sqrt{M^{\;2}_{\rm H} + \vec{q}^{\;2}}$
and $\vec{q}$ are the total energy and the momentum of hydrogen,
$M_{\rm H} = m_p + m_e + \epsilon_{ns}$ and $\epsilon_{ns}$ are the
mass and the binding energy of hydrogen ${\rm H}$ in the $(ns)_F$
hyperfine state; $\phi_{ns}(\vec{k}\,)$ is the wave function of the
$ns$--state in the momentum representation \cite{BS57} (see also
\cite{IV2}--\cite{IV6}). For the calculation of the bound state
$\beta^-$--decay rate we can neglect the hyperfine splitting of the
energy levels of the $ns$--states.

For the amplitude of the bound-state $\beta^-$--decay we obtain the
following expression
\begin{eqnarray}\label{label4}
\hspace{-0.3in}&&M(n\to {\rm H}^{(ns)} + \tilde{\nu}_e) = G_F
V_{ud}\sqrt{2 m_n 2 E_{\rm H}2 E_{\nu}}\nonumber\\
\hspace{-0.3in}&&\times\int
\frac{d^3k}{(2\pi)^3}\,\phi^*_{ns}\Big(\vec{k} - \frac{m_e}{m_p +
m_e}\,\vec{q}\,\Big)\,\Big\{[\varphi^{\dagger}_p
\varphi_n]\nonumber\\
\hspace{-0.3in}&&\times\,[\varphi^{\dagger}_e\chi_{_{\tilde{\nu}_e}}]
- g_A[
\varphi^{\dagger}_p\vec{\sigma}\,\varphi_n]\cdot
[\varphi^{\dagger}_e\vec{\sigma}\,\chi_{_{\tilde{\nu}_e}}]\Big\}.
\end{eqnarray}
The integral over $\vec{k}$ of the wave function
$\phi^*_{ns}(\vec{k}\,)$ defines the wave function $\psi_{ns}(0)$
in the coordinate representation, equal to $\psi_{ns}(0) =
\sqrt{\alpha^3 m^3_e/n^3 \pi}$, where $m_e$ is the electron mass and
$\alpha = 1/137.036$ is the fine--structure constant.  This gives
\begin{eqnarray}\label{label5}
\hspace{-0.3in}&&M(n\to {\rm H}^{(ns)} + \tilde{\nu}_e) = G_F
V_{ud}\sqrt{2 m_n 2 E_{\rm H}2 E_{\nu}}\nonumber\\
\hspace{-0.3in}&&\times\,\psi^*_{(ns)_F}(0)\,\Big\{[\varphi^{\dagger}_{p}
\varphi_{\sigma_n}][\varphi^{\dagger}_{\sigma_e} \chi_{_{\tilde{\nu}_e}}]\nonumber\\
\hspace{-0.3in}&& - g_A[
\varphi^{\dagger}_p\vec{\sigma}\,\varphi_n]\cdot
[\varphi^{\dagger}_e\vec{\sigma}\,\chi_{_{\tilde{\nu}_e}}]\Big\}.
\end{eqnarray}
The bound-state $\beta^-$--decay rate of the free neutron is
\begin{eqnarray}\label{label6}
\hspace{-0.3in}&&\lambda_{\beta^-_b} = \frac{1}{2m_n}\int
\frac{1}{2}\sum^{\infty}_{n=1}\sum_{\sigma_n,\sigma_p,\sigma_e}\!\!\!\!\!|M(n\to
{\rm H}^{(ns)} + \tilde{\nu}_e)|^2\nonumber\\
\hspace{-0.3in}&&\times(2\pi)^4\delta^{(4)}(k_{\nu} + q -
p)\,\frac{d^3q}{(2\pi)^3 2 E_{\rm H}}\frac{d^3k_{\nu}}{(2\pi)^3 2
E_{\nu}}.
\end{eqnarray}
Since the energy shifts of hyperfine interactions is rather small
compared with the energy differences of hydrogen \cite{HFS1}, we
neglect the hyperfine splitting. Summing over all polarisations of the
proton and the electron we take into account the contributions of the
hyperfine states ($ns)_F$ of hydrogen with $F = 0$ and $F =
1$. Summing up over the {\it principal} quantum number and taking into
account that the antineutrino is polarised parallel to its momentum we
get
\begin{eqnarray}\label{label7}
\hspace{-0.3in}\lambda_{\beta^-_b} &=& (1 + 3 g^2_A)\,\zeta(3)\,
G^2_F|V_{ud}|^2\frac{\alpha^3 m^3_e}{\pi^2}\nonumber\\
&\times&\sqrt{(m_p + m_e)^2 + E^2_{\nu}}\,\frac{E^2_{\nu}}{m_n},
\end{eqnarray}
where $\zeta(3) = 1.202$ is the Riemann function, coming from the
summation over the {\it principal} quantum number $n$, and $E_{\nu}$
is equal to
\begin{eqnarray}\label{label8}
E_{\nu} = Q_{\beta^-_c} = \frac{m^2_n - (m_p + m_e)^2}{2 m_n} =
0.782\,{\rm MeV},
\end{eqnarray}
where $Q_{\beta^-_c}$ is the $Q$--value of the continuum-state
$\beta^-$--decay of the free neutron \cite{PDG06}. 

The theoretical value of the continuum-state $\beta^-$--decay rate of
the free neutron is
\begin{eqnarray}\label{label9}
  \lambda_{\beta^-_c} &=& (1 + 3 g^2_A)\, \frac{G^2_F |V_{ud}|^2}{2
    \pi^3}\,f(Q_{\beta^-_c}, Z = 1) = \nonumber\\ &=&1.131\times
    10^{-3}\,{\rm s}^{-1},
\end{eqnarray}
where the continuum-state $\beta^-$--decay rate of the free neutron is
calculated for the experimental masses of the interacting particles
\cite{PDG06} and the Fermi integral $f(Q_{\beta^-_c}, Z = 1)$ equal to
\begin{eqnarray}\label{label10}
\hspace{-0.3in}&&f(Q_{\beta^-_c}, Z = 1) = \nonumber\\
\hspace{-0.3in}&& = \int^{Q_{\beta^-_c} + m_e}_{m_e} \frac{2\pi\alpha
E^2 (Q_{\beta^-_c} + m_e - E )^2 }{\displaystyle 1 - e^{\textstyle
-\,2\pi \alpha E /\sqrt{E^2 - m^2_e}}}\,dE = \nonumber\\
\hspace{-0.3in}&& = 0.059\,{\rm MeV}^5,
\end{eqnarray}
where we have taken into account the contribution of the Fermi
function \cite{HS66}
\begin{eqnarray}\label{label11}
 \hspace{-0.3in}&& F(Z = 1, E) = \nonumber\\
\hspace{-0.3in}&& =\frac{2\pi\alpha E}{\sqrt{E^2 -
m^2_e}}\,\frac{1}{\displaystyle 1 - e^{\textstyle -\,2\pi \alpha
E/\sqrt{E^2 - m^2_e}}}.
\end{eqnarray}
The theoretical value of the lifetime $\tau_{\beta^-_c} =
1/\lambda_{\beta^-_c} = 884.1\,{\rm s}$ agrees well with the
experimental data $\tau^{\exp}_{\beta^-_c} = 885.7(8)\,{\rm s}$
\cite{PDG06}.

For the ratio $R_{b/c} = \lambda_{\beta^-_b}/ \lambda_{\beta^-_c}$
of the bound and continuum state $\beta^-$--decay rates of the free
neutron we get the following expression
\begin{eqnarray}\label{label12}
 R_{b/c} &=& \zeta(3)\,2\pi\, \frac{\alpha^3m^3_e
 E^2_{\nu}}{m_n}\,\frac{\sqrt{(m_p + m_e)^2 +
 E^2_{\nu}}}{f(Q_{\beta^-_c}, Z = 1)} = \nonumber\\
&=&4.06\times 10^{-6}.
\end{eqnarray}
Our value for the ratio of the decay rates agrees well with the
results obtained in \cite{BS1} (see also \cite{WS1,WS2}): $R_{b/c} =
4.20\times 10^{-6}$.

\subsection{Concluding discussion}

Since our calculations are carried out for pure $V-A$ theory of weak
interactions, our results should make corrections to the experimental
analysis of the contribution of scalar and pseudoscalar weak
interactions of hadrons \cite{STW2,WS1,WS2}. We would like to
emphasize that the continuum-state $\beta^-$--decay rate of the free
neutron is sensitive to the value of the axial--vector constant
$g_A$. The value $\tau_{\beta^-_c} = 884.2\,{\rm s}$ is obtained for
$g_A = 1.3$. For the experimental value $g_A = 1.2695$ the lifetime is
$\tau_{\beta^-_c} = 919.7\,{\rm s}$, agreeing with the experimental
value with an accuracy better than $4\,\%$. However, the axial--vector
constant $g_A$ is cancelled for the ratio $R_{b/c}$, therefore our
prediction for the ratio of the bound- and continuum-state
$\beta^-$--decay rates of the free neutron can be valid with an
accuracy much better than $4\,\%$. Since the factor $(1 + 3g^2_A)$
cancels in the ratio $R_{b/c}$, this has no influence on the value
$R_{b/c} = 4.06\times 10^{-6}$. Apart from the radiative corrections
to the continuum-state $\beta^-$--decay rate of the free neutron,
which are of the same order of magnitude \cite{SB58}, the discrepancy
of about $4\,\%$ can be attributed to the contributions of the scalar
and tensor versions of hadronic weak interactions \cite{STW1} (see
also \cite{BS2}), but it is hardly worth to discuss these
contributions in connection with the bound-state $\beta^-$--decay of
the free neutron.

{\renewcommand{\arraystretch}{1.5}
\renewcommand{\tabcolsep}{0.2cm}
\begin{table}[h]
\begin{tabular}{|l|c|c|c|c|}
\hline $\sigma_n$ & $\sigma_p$ & $\sigma_e$ & $\sigma_{\tilde{\nu}_e}$& $f$\\
\hline $+\frac{1}{2} $ & $+\frac{1}{2} $ & $-\frac{1}{2} $
&$+\frac{1}{2} $ & $1 + g_A$\\
\hline $+\frac{1}{2} $ & $+\frac{1}{2}
$ & $+\frac{1}{2} $ &$+\frac{1}{2} $ & $0$ \\
\hline $+\frac{1}{2} $ &
$-\frac{1}{2} $ & $-\frac{1}{2} $ &$+\frac{1}{2} $ & $0$ \\
\hline
$+\frac{1}{2} $ & $-\frac{1}{2} $ & $+\frac{1}{2} $ &$+\frac{1}{2} $ &
$- 2g_A$ \\
\hline $-\frac{1}{2} $ & $+\frac{1}{2} $ & $-\frac{1}{2}
$ &$+\frac{1}{2} $ & $0$ \\
\hline $-\frac{1}{2} $ & $+\frac{1}{2} $ &
$+\frac{1}{2} $ &$+\frac{1}{2} $ & $0$ \\
\hline $-\frac{1}{2} $ &
$-\frac{1}{2} $ & $-\frac{1}{2} $ &$+\frac{1}{2} $ & $1 - g_A$ \\
\hline $-\frac{1}{2} $ & $-\frac{1}{2} $ & $+\frac{1}{2} $
&$+\frac{1}{2} $ & $0$ \\
\hline
\end{tabular}
\caption{The contributions of different spinorial states of the
interacting particles to the amplitudes of the bound-state
$\beta^-$--decay of the free neutron; $f$ is defined by $f =
[\varphi^{\dagger}_e\chi_{_{\tilde{\nu}_e}}][\varphi^{\dagger}_p
\varphi_n] - g_A
[\varphi^{\dagger}_e\vec{\sigma}\,\chi_{_{\tilde{\nu}_e}}]\cdot [
\varphi^{\dagger}_p\vec{\sigma}\,\varphi_n]$.}
\end{table}

Using the amplitude Eq.(\ref{label4}) we can estimate the relative
probabilities of the $n \to {\rm H} + \tilde{\nu}_e$ decays into the
different hyperfine states of hydrogen. Let $(\lambda_{\beta^-_b})_F$
be the decay rate of the bound-state $\beta^-$--decay into the
hyperfine state $(ns)_F$. The ratios of the decay rates are equal to
\begin{eqnarray}\label{label13}
\hspace{-0.3in}&&R_{F = 1} = \frac{(\lambda_{\beta^-_b})_{F =
 1}}{\lambda_{\beta^-_b}} = \frac{3}{4}\,\frac{(1 - g_A)^2}{1
 + 3 g^2_A} = 0.01,\nonumber\\
\hspace{-0.3in}&&R_{F = 0} =
 \frac{(\lambda_{\beta^-_b})_{F = 0}}{\lambda_{\beta^-_b}} =
 \frac{1}{4}\,\frac{(1 + 3g_A)^2}{1 + 3 g^2_A} = 0.99,
\end{eqnarray}
calculated for both the experimental value of the axial coupling
constant $g_A = 1.2695$ and $g_A = 1.3$ \cite{PDG06}.

This means that in the final state of the $n \to {\rm H} +
\tilde{\nu}_e$ decay hydrogen is produced in the hyperfine state with
$F = 0$ with a probability $99\,\%$. The main part of this probability
$83.36\,\%$ is caused by the transition to the ground hyperfine state
$(1s)_{F = 0}$.

We are grateful to Prof. T. Ericson for calling our attention to the
problem of the bound-state $\beta^-$--decay of the free neutron and
the proposal to analyse this problem in our approach to the
$\beta$--decays of highly charged heavy ions \cite{Ivanov1,Faber1} and
mesic hydrogen \cite{Faber2}.

\end{document}